\documentclass[
reprint,
superscriptaddress,
amsmath,
amssymb,
aps,
prb,
longbibliography,
noeprint
]{revtex4-2}

\usepackage{epsfig}
\usepackage{dcolumn}
\usepackage{bm}
\usepackage{color}
\usepackage{graphicx}
\usepackage{bbold}

\begin{document}

\title{Microscopic scale of quantum phase transitions: from doped semiconductors to spin chains, cold gases and moiré superlattices}

\author{Andrey Rogachev} 
\affiliation {Department of Physics and Astronomy, University of Utah, Salt Lake City 84093, USA}

\date{\today}

\begin{abstract}
In the vicinity of continuous quantum phase transitions (QPTs), quantum systems become scale-invariant and can be grouped into universality classes characterized by sets of critical exponents. We have found that despite scale-invariance and universality, the experimental data still contain information related to the microscopic processes and scales governing QPTs. We have found that for many systems, the scaled data near QPTs can be approximated by the generic exponential dependence introduced in the scaling theory of localization; this dependence includes as a parameter a microscopic seeding scale of the renormalization group, $L_0$. We have also conjectured that for interacting systems, the temperature cuts the renormalization group flow at the length travelled by a system-specific elementary excitation over the life-time set by the Planckian time, $\tau_P$=$\hbar/k_BT$. We have adapted this approach for QPTs in several systems and showed that $L_0$ extracted from experiment is comparable to physically-expected minimal length scales, namely (i) the mean free path for metal-insulator transition in doped semiconductor Si:B, (ii) the distance between spins in Heisenberg and Ising chains, (iii) the period of an optical lattice for cold atom boson gases, and (iv) the period of a moiré superlattice for the Mott QPT in dichalcogenide bilayers. The metal-insulator transition in Si:P has been explained using a non-interacting version of the model. In two companion papers, we show that in superconducting systems, $L_0$ is comparable to superconducting coherence length, and in quantum Hall systems, to the magnetic length. The developed new method of data analysis identifies microscopic processes leading to QPTs and quantitatively explains and unifies a large body of experimental data.

\end{abstract}

\maketitle

\noindent \textbf{1. Introduction.}

	A quantum phase transition (QPT) is a transformation between ground states of a quantum system driven by a non-thermal parameter, $y$, such as a magnetic field or chemical potential \cite{Sachdev_Book}. QPTs appear in a large variety of systems in nature ranging from magnetic materials \cite{Lohneysen_Review,Zheludev_Review} and semiconductors \cite{Bogdanovich_SiB,Lohneysen_SiP} to cold atoms \cite{Chin_2d_ColdAtoms,Yang_1d_ColdAtoms}, atomic nuclei \cite{Cejnar_QPT_nuclei} and stars \cite{Alford_QPT_quark}. More recently, this list has grown to include QPTs in graphene and dichalcogenide moiré superlattices \cite{Yankowitz_QPT_BiGraphene,Mak_QPT_Mott} as well as in topological insulators \cite{Kawamura_anomQHE,Liu_QPT_Chern}. 
  
	Strictly speaking, QPTs occur at zero temperature when the driving parameter $y$ reaches its critical value $y_c$. They are marked by a diverging spatial correlation length of quantum fluctuations, $\xi\sim |y-y_c|^{-\nu}$, where $\nu$ is the correlation length critical exponent. The dynamics of the fluctuations is characterized by a temporal scale $\xi_\tau$, which is related to $\xi$ by the dynamical critical exponent $z$, as  $\xi_\tau\sim\xi^z$. Physics of QPTs treats space and time scales equally; this often allows to map a d-dimensional quantum system on a classical system with an effective dimension $d_{eff}$=$d+z$ \cite{Hertz,Sondhi_QPT}. 

	According to the theories of QPT, the temperature sets a finite size of a system in the (imaginary) time direction by a quantity $\tau_P\sim\hbar/k_BT$ \cite{Sachdev_Proceedings,Damle_Sachdev,Sondhi_QPT}. This quantity is termed as the Planckian dissipation time \cite{Phillips_StrangeMetal,Hartnoll_Planckian}.  In interacting systems near a QPT, the Planckian time sets the shortest equilibration time of local excitations \cite{Sachdev_Keimer}. Further, the theory introduces the spatial dephasing length related to temperature as $L_\varphi\propto T^{-1/z}$.

\begin{figure}[b]
\centering
 \includegraphics[width= 0.7\columnwidth]{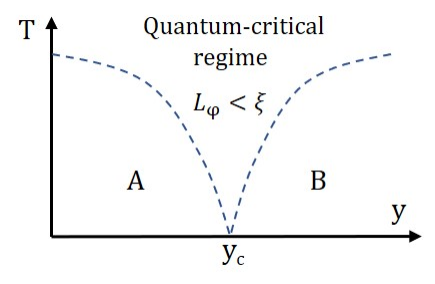}
 \caption{Generic diagram of a quantum phase transition driven by parameter $y$. 
 Regions A and B corresponds to two phases undergoing a QPT. In the quantum critical regime, the dephasing length $L_\varphi$ is smaller than the correlation length $\xi$.}
 \end{figure}
Two length scales, $\xi$ and $L_\varphi$,  define a phase diagram of a QPT, as shown in Fig. 1. The regions A and B are dominated by the properties of the phases undergoing a QPT. For example, these could be ordered and paramagnetic states in a magnetic system \cite{Kirnoss_1d_Ising} or metal and insulating states in a localization-delocalization transition \cite{Gantmakher_LocDeloc}. In the quantum critical regime, the temperature $T$ sets the energy scale. The boundaries of this regime are determined by the condition  $L_\varphi(T)<\xi$ \cite{Gantmakher_LocDeloc}, though other conditions are also used \cite{Sachdev_Book}. The physical meaning of this is that the temperature breaks the system into uncorrelated volumes on a scale smaller than $\xi$. 

The wave function of a system in the quantum critical regime has a strongly entangled form. An accurate theoretical description of this regime is a major challenge and apparently has not been achieved thus far \cite{Sachdev_Keimer}; the complexity of real materials also further obscures the analysis.  The exceptions to this have been several one-dimensional experimentally testable systems which allow for a complete quantitative theoretical description, starting from microscopic processes governing critical fluctuations and going all the way to the prediction of the long-range behavior near the critical point.   These systems are mostly limited to materials with one-dimensional (1d) spin chains and ladders \cite{Zheludev_Review}, with the recent addition of 1d superconducting nanowires \cite{Rogachev_QPR_wires} and 1d cold gases \cite{Yang_1d_ColdAtoms}.  There appears to be only one example of a system in higher dimensions which matches this level of understanding, the 3d insulating magnet LiHoF$_4$ \cite{Bitko_LiHoF3}.

The vast majority of experimental data on QPTs have been analyzed using the phenomenological finite-size scaling theory \cite{Sondhi_QPT}. This theory predicts that in the quantum critical regime, many physical quantities are described by corresponding scaling equations.  For example, the zero-bias DC electrical conductivity takes the form
\begin{equation}
\sigma(y,T)=\frac{e^2}{\hbar} L_{\varphi}(T)^{-(d-2)} \Phi_\sigma\left(\frac{y-y_c}{T^{1/z\nu}}\right).
\end{equation}
Here $d$ is dimensionality and $\Phi_\sigma$ is the scaling function. In most of the cases, the scaling functions are unknown and, in the analysis, one varies the critical exponents and $y_c$  to obtain the best “scaling collapse” of the data. 
	
A major deficiency of such analysis is that it does not disclose the microscopic mechanism of the transition. Extracted values of exponents are often compared to rather simplistic models, the applicability of which to a particular system is not always certain. A good example is the Chalker-Coddington network model and its generalizations, which were used to explain QPT transitions in integer and fractional Quantum Hall systems \cite{Chalker,Huckestein,Jain_fQHE_theory,Lee_Wang_Kivelson}. These models accurately predict observed critical exponents but completely ignores Coulomb interactions present in 2d electron-gas; a rather dramatic view on this discrepancy is given in \cite{Werner_Oswald_QHE}.
 
	 Another long-standing problem is the metal-insulator transition (MIT) in 2d electron gas. Despite an early observation of excellent scaling of the data \cite{Kravchenko_Scaling} and significant research efforts \cite{Spivak_2dMIT_review,Shashkin_Review}, the mechanism of the transition is still not understood. The extracted values of exponents are system- and sample-dependent \cite{Pudalov_2dMIT_scaling}. There are several proposed theories for this QPT and alternative interpretations stating that the experimental $R(T)$ curves only mimic QPT behavior and in reality, have a different physical origin (see reviews \cite{Spivak_2dMIT_review,Shashkin_Review} for details).  

As with the MIT in 2d electron gas, the critical exponents for the superconductor-insulator transitions (SIT) in 2d systems appear to vary unsystematically \cite{Sacepe_review}. Moreover, in our recent study, we employed an intentionally dimensionality-inconsistent analysis of 1d superconducting nanowires using Eq. 1 in the form suitable to 2d films and found an excellent but completely accidental “data collapse” \cite{Rogachev_Deficiency}. This shows that the scaling of data by itself is not a particularly reliable indicator of QPT.
  
	In this paper, we present a phenomenological model which goes beyond the standard finite-size scaling and gives an access to the microscopic physics of QPTs. Similar to classical phase transitions, QPTs should follow the renormalization group approach.  It is generally believed that in the process of renormalization and coarse-graining, information about microscopic nature of a system is lost and only its most universal characteristics are preserved. We have found that the situation is not that restrictive in the quantum critical regime and that a microscopic seeding scale, $L_0$, of the renormalization groups can, in fact, be extracted from experimental data. The strength of this analysis comes from the fact that, for many systems, it is fairly clear what the seeding scale is and so experimental verification becomes possible.
  
  A creative step of the procedure is to figure out the system-dependent form for $L_\varphi$. We have discovered that, for interacting systems, agreement with experiment occurs when we equate $L_\varphi$ to a distance traveled by a non-interacting particle or elementary excitation during time  $\tau_P$=$\hbar/k_BT$.  
  
  The details of the model are presented in the next section and in consequent sections, we use it to analyze QPTs in several classes of systems using data traced from literature. In two companion papers, we show the model can be applied to superconducting films and nanowires as well as to 2d topological systems \cite{Rogachev_QPT_films,Rogachev_QPT_QHE}. The model is based on several conjectures, theoretical justifications for which are incomplete at this point. 
\smallbreak
\noindent \textbf{2. Model}

Our work started with the observation that the scaled conductivity in superconducting MoGe films displays a roughly exponential variation across the magnetic-field-driven SIT. (See Fig.2 in \cite{Rogachev_QPT_films}).  Extensive inspection of the literature has indicated that this variation appears in many other systems, including magnetic and cold atom systems. This similarity between very distinct systems has largely gone unnoticed or unreported likely due to the practice of presenting data on a log-log scale as a function of the absolute value of the argument of the scaling function. Plotted in this way, data displays a “mirror” symmetry which is often discussed in terms of duality between two phases undergoing QPT.  

For MoGe films, where we are certain that the QPT is of a pair-breaking type and for many other systems, the duality picture is not appropriate.  The needed exponential variation, however, comes from the functional form chosen approximate the $\beta$-function across the metal-insulator transition for disordered 3d systems in the scaling theory of localization (STL) \cite{STL}. We proceeded with an analysis guided by a conjecture that this approximation the essential ingredients of the real space one-parameter renormalization group and, in this regard, can be extended to many other systems.      

\begin{figure}[t]
\centering
 \includegraphics[width= 0.7\columnwidth]{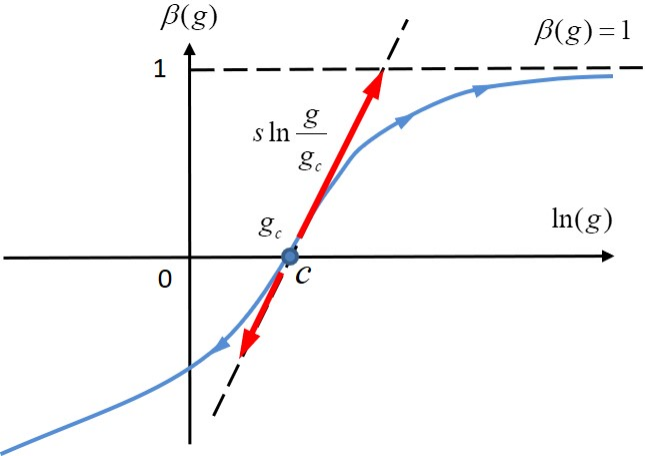}
 \caption{Variation of the $\beta$ function (blue line) versus logarithm of the dimensionless conductance $g$ suggested in the scaling theory of localization for 3d disordered non-interacting system. At the critical point $C$ the conductance is $g_c$. The red arrows indicate flow of the system with increasing size along which the integration of the $\beta$-function is taken.}
 \end{figure}

The scaling theory of localization introduces the dimensionless conductance, $g$, of a hypercube of size $L^d$, which is related to the conductance, $G$, and conductivity, $\sigma$, as $\sigma$=$G/L^{d-2}$=$(e^2/\hbar)(g/L^{d-2})$. The theory assumes that the conductance of a cube of a bigger size $b^dL^d$ is determined only by $b$ and $g(L)$. In continuous form, this statement is expressed as the scaling equation for the function $\beta$ where $\beta(g(L))$=$d\ln {g(L)}/d \ln{L}$. This function describes how the conductance changes or “flows”, using the language of the renormalization group, with increasing system size. The flow starts at some microscopic scale, $L_0$, with a dimensionless conductance, $g_0$. Figure 2 illustrates this process. 

Near the critical point of the metal-insulator transition (MIT), that is, near the critical conductance $g_c$,  $\beta$ behaves approximately linearly as $\beta$=$s\ln{g/g_c}$.  This approximation, though not exactly proven, was used in the original paper on SLT \cite{STL} as well as in some reviews \cite{Gantmakher_LocDeloc} and textbooks \cite{Imry_Book,Girvin_Book}.  In the appendix, we give a brief review of STL and show that the coefficient, $s$, in this equation is equal to $1/\nu$. 

The first steps in our analysis are similar to a model proposed in \cite{Dobro_2dMIT_STL} to explain the exponential variation of conductance across MIT in 2d semiconductors. Using the linear approximation for $\beta$, we integrate the general scaling equation for $\beta(g(L))$ starting from some microscopic conductance, $g_0$, corresponding to some microscopic length scale, $L_0$. The integration gives $\ln (g/g_c)$=$\ln (g_0/g_c)(L/L_0)^{1/\nu}$. Exponentiating both sides and converting the equation to the conductivity of a cube with side $L$, we find that $\sigma$=$e^2 g_c\exp(\ln(g_0/g_c)(L/L_0)^{1/\nu})/(\hbar L^{d-2})$.  

Then, using the expansion $\ln{\left(g_0/g_c\right)}\approx\left(g_0-g_c\right)/g_c$ and an approximation that near the critical conductance, $g_c$, the conductance changes linearly with the driving parameter $y$ as $\left(g_0-g_c\right)/g_c\approx\left(y-y_c\right)/y_c$ we obtain 
\begin{equation}
\sigma = \frac{e^2}{\hbar L^{d-2}} g_c \exp \left(\frac{y-y_c}{y_c} \left(\frac{L}{L_0}\right)^{1/\nu} \right)
\end{equation}
Equation 2 describes the variation of the zero-temperature conductivity of a system as a function of its size $L$. At finite temperatures, in the quantum critical regime, thermal fluctuations break the system coherence, so the variation given by Eq.2 is cut and switches into the ohmic regime at the dephasing length as determined by the temperature and the dynamical exponent, $z$, as $L_\varphi \propto T^{-1/z}$. Mathematically this means that  $L$ in Eq. 2 is replaced by  $L_\varphi$, 
\begin{equation}\sqrt{}
\sigma(T)=\frac{e^2}{\hbar L_\varphi^{d-2}}g_c\exp \left(\frac{y-y_c}{y_c}\left(\frac{L_\varphi(T)}{L_0}\right)^{1/\nu}\right)
\end{equation}
In the appendix we show that the scaling theory of localization predicts that the zero-temperature spatial correlation length diverges at the transition as  $\xi\approx L_0(y_c/|y-y_c|)^\nu$. The absolute value of the exponent argument in Eq.3 can be expressed as  $(L_\varphi(T)/\xi)^{1/\nu}$ and, hence, the quantum critical regime is limited by about one order of magnitude in the exponent variation (for example, for $\nu=1$, the change is between $e^{-1}$ and $e^1$). 
 The model was originally developed to explain our data for the QPT in superconducting films \cite{Rogachev_QPT_films}. As we show below, it actually has a much broader range of applications.

\smallbreak
\noindent \textbf{3. Metal-Insulator transition in 3d disordered materials. }

\textit{Si:P doped crystalline semiconductor.}

\begin{figure*}
\centering
 \includegraphics[width= 0.8\textwidth]{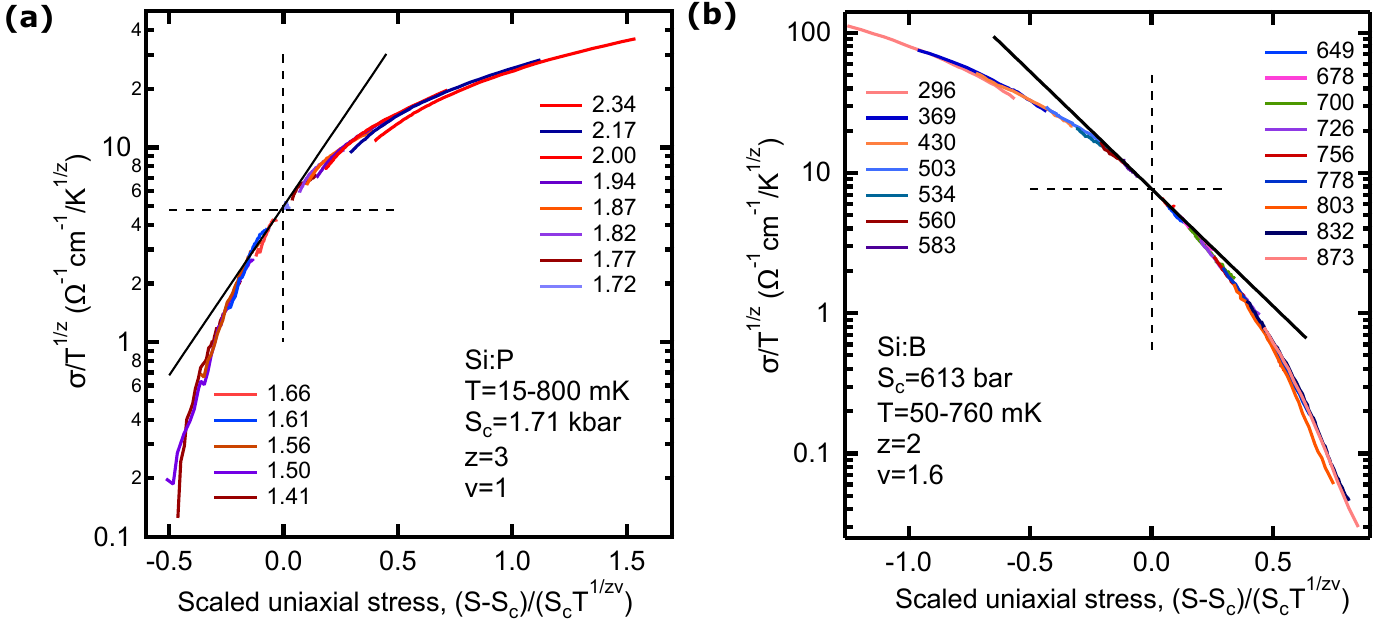}
 \caption{\textbf {Metal-insulator transition in 3d disordered materials.} (a) Scaled conductivity versus scaled uniaxial stress for crystalline Si:P studied in \cite{Lohneysen_SiP}. The stress values for each experimental curve are indicated. 
 (b) Scaled conductivity versus scaled uniaxial stress for crystalline Si:B studied in \cite{Bogdanovich_SiB}. The stress values for each experimental curve are indicated. In both panels, the solid lines shows the exponential fit near the critical point.}
\end{figure*}

We start our analysis with the metal-insulator transition (MIT) driven by the uniaxial stress, $S$, in phosphorous-doped crystalline silicon, Si:P, as reported in \cite{Lohneysen_SiP}. We have traced the data from Fig. 1 in \cite{Lohneysen_SiP} and used the values of the critical exponents found in this work, $\nu\approx1$ and $z\approx3$.  The scaling plot is shown in Fig. 3a with axis suitable for our analysis. The solid black line gives the exponential approximation of the data near the critical point  $y$=$a_1 \exp (a_2x)$,  with $a_1$=5.0 $\Omega^{-1}$cm$^{-1}$K$^{-1/3}$ and $a_2$=4.0 K$^{-1/3}$.  
	
The value of the critical exponent, $z$=$d$=3,  suggests that Si:P undergoes a MIT as a \textit{non-interacting} system \cite{Girvin_Book} (page 289). Same reference also suggests that in a non-interacting 3d system at zero temperature, the characteristic time scale relates to length as $\tau_L\sim h \frac{dn}{d\mu}L^3$, where $dn/d\mu$ is the electron compressibility.  Extending this reasoning, we argue that at finite temperatures in the quantum critical regime, the time scale is set by temperature as $h/\tau_L \sim k_B T$ and the length scale by the dephasing length defined as $L_\varphi \approx (\frac{dn}{d\mu} k_BT)^{-1/3}$. The approximation for the scaling function near the critical point then becomes 
\begin{equation}
\sigma=\frac{g_c e^2}{\hbar}\left[\frac{dn} {d\mu} k_BT\right]^{1/3} \exp \left( \left[ \frac{dn}{d\mu} k_B \right]^{-1/3} \frac{1}{L_0}\frac{S-S_c}{S_c T^{1/3}} \right) 
\end{equation}
Compared to the generic Eq. 1, Eq. 4 has a defined scaling function with a dimensionless argument and two additional parameters $g_c$ and $L_0$, which can now be extracted from experiment and used to test the assumptions of the model. 

To proceed with the analysis, we need to find $dn/d\mu$. It is a non-trivial question. In disordered systems, electron-electron interactions lead to the formation of the zero-bias anomaly in the single-particle density of states measured by tunneling. This anomaly transforms into a Coulomb gap on the insulating side of the transition. The theories of MIT, however, tell us that $dn/d\mu$ is not affected by electron-electron interactions and goes smoothly across the transition. This non-critical variation is also a property of the density of states $g(E_F)$ associated with specific heat coefficient, $\gamma$. (See, however, a recent report on anomalous downturn of $\gamma$ in Ti-Si amorphous alloys at low temperatures \cite{Rogachev_SH_critical}). In the non-interacting Fermi gas, $g(E_F)$ and $dn/d\mu$ are equal to each other. Therefore, we believe that $g(E_F)$ can be used as an approximation for $dn/d\mu$ in Eq. 4. 

The studied Si:P sample has a phosphorus concentration of $n_P$=$3.21\times{10}^{18}$ cm$^{-3}$. This gives the estimate for the average distance between phosphorous atoms and, correspondingly, the elastic mean-free path as $\ell\approx a\approx n_P^{-1/3}$=6.8 nm. 
Measurements of the specific heat of Si:P report a value of $\gamma\approx 30$  $\mu$J/mole K$^2$ for a sample with $n_P\approx 3\times 10^{18}$ cm$^{-3}$ \cite{Lohneysen_SiP_SH}. Using the formula $\gamma$=$\pi^2 k_B^2 g(E_F)/3$, we estimate the density of states as $g(E_F) \approx 6.4 \times 10^{20}$ eV$^{-1}$cm$^{-3}$. The seeding scale then can be estimated as $L_0\approx a_2^{-1}(g\left(E_F\right)k_B)^{-1/3}$=7.8 nm.
	 
We are now able to see that $L_0$, the microscopic seeding scale of MIT in Si:P, is on the order of the mean free path. In strongly disordered materials, the conductance is not defined at length scales below $\ell$, so the relation $L_0\approx\ell$ is indeed expected as was suggested in the original paper on the scaling theory of localization. 

The experimental value of the critical conductance is $g_c$=0.06. Note that this value depends on the choice of units used to define dimensionless conductance. If, following \cite{Girvin_Book} (page 280), we use relation $G=2e^2/h\times g$, we obtain $g_c \approx 0.2$, which is closer to the value $g_c\approx1$, suggested in STL.

\textit{Si:B doped crystalline semiconductor.}

Let us now analyze MIT in crystalline silicon doped with boron, Si:B, also driven by uniaxial stress. The data were traced from Fig. 1 in \cite{Bogdanovich_SiB}.  The scaling analysis was performed in this work, from which the values of critical stress and critical exponents, $\nu$=1.6 and $z$=2 , were determined. We plot the scaled data in Fig.3b, again in a form suitable for comparison with our model. 

The experimental value of the dynamical critical exponent of $z$=2 and the disordered nature of the material suggest a simple relation for the dephasing length, $L_\varphi$=$\left(b_z\tau_P\right)^{1/z}\approx\left(\hbar D/k_BT\right)^{1/2}$.  The value of the exponent implies that we are dealing with an interacting system (page 289 in \cite{Girvin_Book}). 

To move forward, we make here our \textit{second conjecture}: we assume that all interaction effects are incorporated into $\tau_P$ and that the size of the fluctuating volume, as determined by $L_\varphi$, is simply the distance traveled by a non-interacting particle over this time. For the specific example of Si:B, this means that the diffusion coefficient, $D$, is the Drude diffusion coefficient and does not include quantum corrections or complications due to proximity to MIT. Support for this non-critical behavior also comes from dimensional arguments (page 290 in \cite{Girvin_Book}), which suggest that in interacting systems, the diffusion coefficient should scale as $D \sim L^{2-z}$ and thus remains constant in systems with $z$=2. 

Returning to Eq. 3 with $d$=3,  $\nu$=1.6, $z$=2, and $L_\varphi$=$\left(\hbar D/k_BT\right)^{1/2}$, we have for Si:B:
\begin{equation}
\sigma=\frac{g_c e^2}{\hbar}\left(\frac{k_B T} {\hbar D}\right)^{1/2} \exp \left(- \left[ \frac{\hbar D}{k_B} \right]^{1/3.2} \frac{1}{L_0^{1/1.6}} \frac{S-S_c}{S_c T^{1/3.2}} \right) 
\end{equation}
The minus sign in the exponent appears because the conductivity decreases with increasing stress.  We approximate the data near the critical stress value, $S_c$, with an exponential function $y$=$a_1 \exp(-a_2x)$, where $a_1$=7.7 $\Omega^{-1}$ cm$^{-1}$K$^{-1/2}$ and $a_2$=3.85 K$^{-1/3.2}$; this is shown in Fig. 3(b) as a solid line.  To find the experimental $g_c$ and $L_0$ we now need to estimate the diffusion coefficient, $D$.

	The Si:B sample tested in \cite{Bogdanovich_SiB} has dopant concentration $n$=4.84$\times{10}^{18}$ cm$^{-3}$, for which the average distance between the boron atoms is $a\approx n^{-1/3}$=5.9 nm. For elastic mean free path we have $\ell\approx a$.  We estimate the density of state effective mass, $m_{dh}$=0.55$m$,  using formula $m_{dh}^{3/2}$=$m_{hh}^{3/2}+m_{hl}^{3/2}$; $m_{hh}$=0.49$m$ and $m_{hl}$=0.16$m$ are the effective masses for heavy and light holes in Si, respectively, and $m$ is the free electron mass \cite{Sze_Book}. This gives the Fermi energy, $E_F\approx0.019$ eV, and the density of states, $N(E_F)\approx 4.1 \times{10}^{20}$ eV$^{-1}$cm$^{-3}$.  Estimating the conductivity mass as $m_c\approx\left(m_{hh}m_{hl}\right)^{1/2}$=0.28$m$, we find the Fermi velocity, $v_F\approx1.54\times{10}^5$ m/s, and the diffusion coefficient, $D$=$v_F\ell/3\approx3.0$ cm$^2$/s.  Using the experimental values of the coefficients $a_1$ and $a_2$, we then find $L_0\approx5.6$ nm and $g_c\approx0.15$ and again arrive at physically significant result that $L_0\approx\ell$. 	

The two examples presented in this section demonstrate that the analysis of the quantum critical regime can be done quantitatively and extended beyond the determination of the critical exponents. While the extremely close agreement between  $\ell$ and $L_0$ in doped Si looks fortuitous, it is not accidental. If our model is wrong, there is no reason for these two quantities not to be off by many orders of magnitude. 

Our model is phenomenological and obviously leaves many questions unanswered. It indicates that the MIT in Si:P is due to localization, but does not explain why the correlation length exponent takes the value  $\nu$=1, as predicted by the nonlinear-$\sigma$ model \cite{Hikami_nonliner_sigma} and self-consistent diagrammatic theory \cite{Shapiro_MIT}, and not $\nu$=1.5  as predicted by numerical simulations \cite{Kramer_MacKinnon}.  It is also not clear why, in sharp contrast to Si:P, the transition in Si:B goes via the interacting mechanism. Is it because in doped semiconductors, the Anderson transition appears roughly at the same carrier concentration as the Mott transition and small distinctions can push a system toward one or another route? Or does the critical state in Si:B emerge because of the interaction between many carriers? Hopefully some of these questions can be addressed by making a comparison with clean systems, which we consider in the next sections.  

\smallbreak
\noindent \textbf{4. Quantum phase transition in spin chains. }

An initial incentive to apply our model to spin-chain materials came from the observation of a roughly exponential variation of the nuclear spin-lattice relaxation rate $t_1^{-1}$ reported for 1d spin-chain material NiCl$_2$-4SC(NH$_2$)$_2$ (DNT) (Fig. 3 in \cite{Mukhopadhyay_1d_Heisenberg}).  In spin-chains and ladders, spins are positioned periodically in space and these systems are technically clean. However, close to a QPT, they can be thought of as a mixture of two phases and thus are in a strongly disordered state. The main procedure of the scaling theory of localization, which considers the system evolution according to the boundary condition of stacking blocks of increasing size, then appears as a plausible way of thinking about QPT in these systems.  

\begin{figure*}
\centering
 \includegraphics[width= 0.8\textwidth]{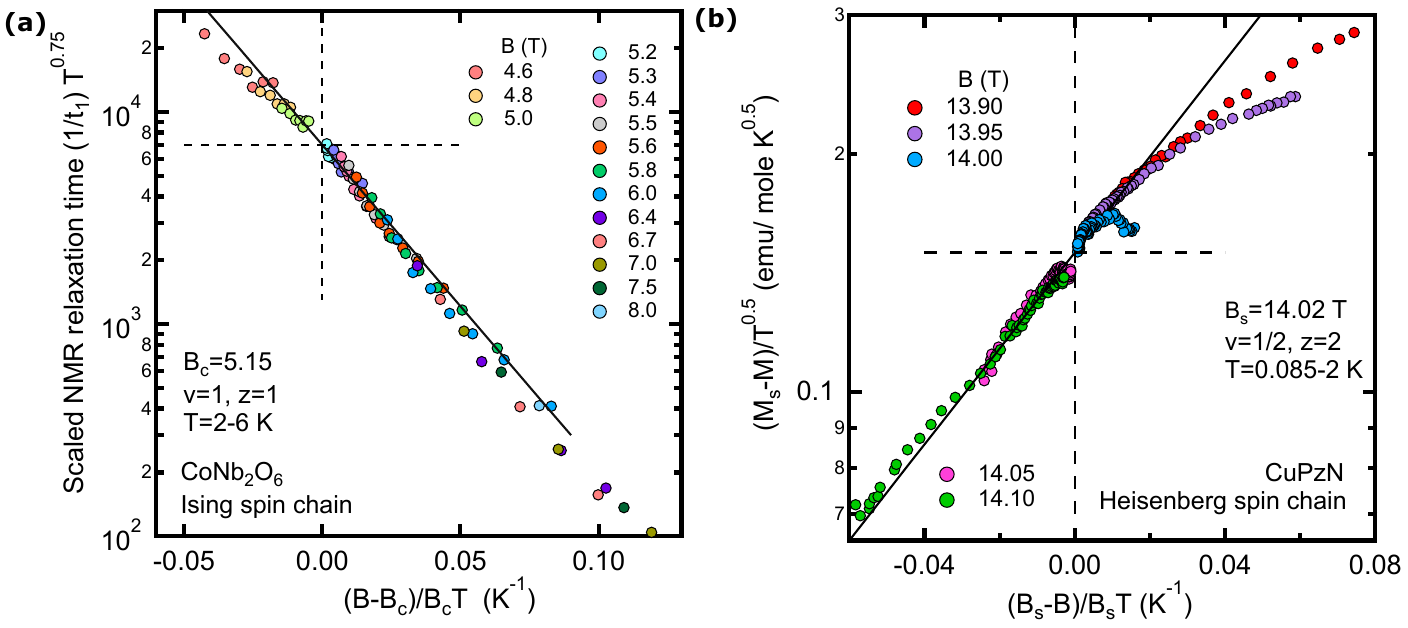}
 \caption{\textbf{QPT in magnetic materials with Ising and Heisenberg spin chains.} (a) Scaled NMR relaxation rate $1/t_1$ plotted versus scaled magnetic field for Ising spin chain material CoNb$_2$O$_6$ studied in \cite{Kirnoss_1d_Ising} (b) Scaled magnetization versus scaled magnetic field for Heisenberg spin chain material CuPzN studied in \cite{Kono_1d_Heisenberg}.}
 \end{figure*}

\noindent \textit{4.1 1d Ising spin chains}
	
We first analyze QPTs in the ferromagnetic 1d Ising chain material Co$_2$Nb$_2$O$_6$; the transition is driven by a magnetic field applied perpendicular to the direction of the spins. The evolution of the nuclear spin-relaxation time, $t_1$, across the transition was reported for this system in \cite{Kirnoss_1d_Ising}; we have traced the raw data from Fig.  5 in this work. The scaling analysis carried out in \cite{Kirnoss_1d_Ising} differs from ours as the authors apparently used it for the renormalized classical and quantum disordered regimes (A and B sections in Fig. 1). 

Our goal is the quantum critical regime. Accordingly, we multiplied the relaxation rate by $T^{3/4}$ to accommodate the dependence $1/t_1 \sim T^{-3/4}$ theoretically computed in \cite{Kirnoss_1d_Ising} for the critical field $B$=$B_c$ and then scaled the data. We found that the best collapse occurs at $\nu z$=1 and $B_c$=5.15 T. The latter value is somewhat smaller than $B_c$=5.3 T  reported in \cite{Kirnoss_1d_Ising}.  The resulting scaling plot is presented in Fig. 4(a). The scaled data display roughly exponential variation across the transition; the solid line shows the fit near $B_c$ to $y$=$a_1 \exp (-a_2x)$, with  $a_1$=7000 s$^{-1}$K$^{3/4}$ and $a_2$=35  K$^{-1}$.

Let us now adapt Eq. 3 to the QPT in Ising chains materials.  At the critical point of the transition, the speed of elementary excitations, domain walls in the ferromagnetic state, becomes $c$=$2Ja/ \hbar$ \cite{Sachdev_Book}, where $J$  is the exchange coupling and $a$ is the distance between the spins in the chains. In the previous section, we made the conjecture that in the quantum critical regime, the dephasing length is set by the distance traveled by an elementary excitation during the Planckian time $\tau_P$. For the Ising spin chain, this is just the $\textit{ballistic}$ distance  $L_\varphi\approx c\tau_P$=$(2Ja/\hbar)(\hbar/k_BT)$.  This expression implies the dynamical exponent $z$=1, in agreement with the prediction of the microscopic critical theory \cite{Sachdev_Book}.  The correlation length exponent extracted from the fit, $\nu$=1, also agrees with the theoretical prediction.  

Within the scaling theory of localization, Eq. 3 explains the variation of the electrical conductivity in disordered systems. To expand its applicability to other observables and other systems, we assume that only the exponential part of the equation is generic for the model.  The behavior of the prefactor depends on more specific details of a particular study. For the present case, the prefactor takes the form $AT^{-3/4}$, where $A$ is the constant. The resulting scaling equation is 
\begin{equation}
t_1^{-1}=AT^{-3/4} \exp\left(-\frac{2Ja}{k_BL_0}\frac{B-B_c}{B_cT}\right). 
\end{equation}
Using the value of the super-exchange interaction between Co ions determined in the neutron scattering experiments \cite{Coldea_E8},  $J$=1.94 meV, and the experimental value of the coefficient $a_2$, we find that the seeding scale of the transition, $L_0$=1.3$a$, is very close to the distance between the spins $a$. This is indeed the quantity expected to represent the minimal seeding scale. 
\break
 	
\noindent \textit{4.2 1d Heisenberg spin chains}

Let us now turn to the second class of 1d spin systems, materials which can be mapped onto Heisenberg antiferromagnetic spin-1/2 chains. First, we analyzed magnetization data for Cu(C$_4$H$_4$N$_2$)(NO$_3$)$_2$, or CuPzN for short, reported in \cite{Kono_1d_Heisenberg}, which according to this work is a practically perfect spin-1/2 Heisenberg antiferromagnet. With increasing magnetic field, CuPzN undergoes a QPT from a Tomonaga-Luttinger liquid (TLL) to a field-induced ferromagnetic state.  Near the critical field $B_s$, an effective description of the TTL can be given in terms of interacting magnons, which in the dilute limit can be exactly mapped onto free fermions with a chemical potential defined as $\mu$=$g\mu_B(B_s-B)$. The ground state at $B>B_s$ can be considered as a vacuum state. In the quantum critical regime, predicted exponents are $\nu$=1/2  and $z$=2. (References to the theoretical works are given in \cite{Kono_1d_Heisenberg}). The magnetization in CuPzN was found to behave according to the these predictions; in particular,its with field at $T$=0.08 K was accurately matched by the exact quantum transfer matrix calculations.   

Despite the fact that the magnetism in CuPzN is a solved case, we want to see if our model provides a complimentary view of its quantum critical regime.  We traced the experimental data from Figs. 2 and 3 of \cite{Kono_1d_Heisenberg} and scaled them; our scaling plot is shown in Fig. 4(b). In this plot, $M$ is the magnetization and $M_s$ is its saturation value at high magnetic fields; the only parameter we adjusted was $B_s$. The product $\nu z$=1 and the power of the temperature in the prefactor, $T^{0.5}$, were preset to the theoretical values. The scaling collapse appears to be good. The deviation for the data at $B$=14.00 T represents a fairly common situation in which a noise and systematic non-critical contribution are excessively magnified for traces close to the critical point.  The solid line shows the exponential fit to the data, $y$=$a_1\exp (a_2x)$, with  $a_1$=0.15 emu/mole K$^{0.5}$ and $a_2$=14 K$^{-1}$.   

\begin{figure*}
\centering
 \includegraphics[width= 0.8\textwidth]{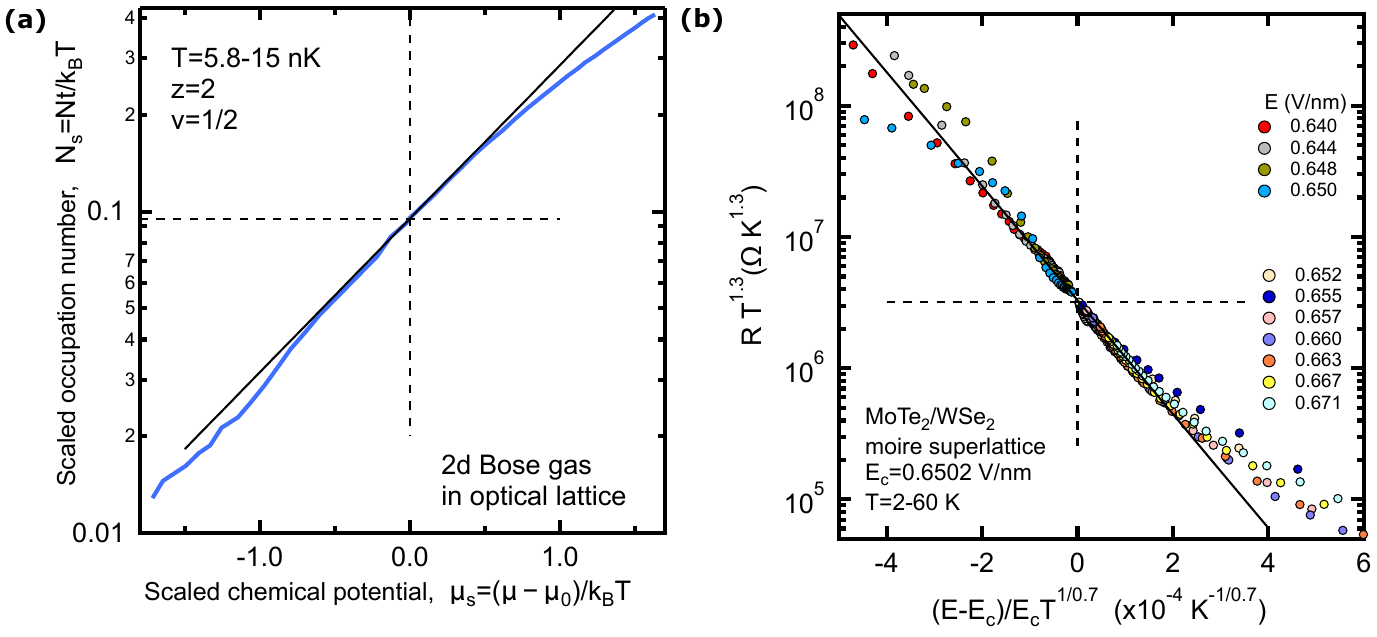}
 \caption{\textbf{QPT in optical and moiré lattices.} (a) Scaled occupation number versus scaled chemical potential for cold gas of Cs-133 atoms in two-dimensional optical lattice studied in \cite {Chin_2d_ColdAtoms}. (b) Scaled resistance per square versus scaled electrical field for MoTe$_2$/WSe$_2$ moiré superlattice studied in \cite{Mak_QPT_Mott}. }
 \end{figure*}
	
To proceed with the analysis, we need to find the dependence of $L_\varphi$ on temperature. We have essentially guessed it guided by the physics of the system and the expectation that we need to reproduce the $z$=2 exponent predicted theoretically. At $B$=$B_s$ and zero temperature, the chemical potential is zero and the system is in the vacuum state with no magnons excited above the ferromagnetic ground state.  At finite temperatures, the elementary excitations, magnons, behave as Maxwellian free particles with typical velocity set by the temperature, $v$=$\sqrt{k_BT/m_m}$. The motion of these particles is ballistic, so during the Plankian time, they travel the distance, $L_\varphi$=$v\tau_P$=$\hbar\sqrt{1/k_BTm_m}$.  Let’s notice that up to a coefficient of order of one, this is also the de Broglie wavelength of the magnons. The effective mass of a magnon is $m_m$=$\hbar^2/Ja^2$, where $a$ is the lattice constant and $J$=10.8 K is the intrachain coupling constant for CuPzN \cite{Kono_1d_Heisenberg}. With these parameters and $\nu$=1/2, the scaling equation for a spin-1/2 chain becomes
\begin{equation}
M_s-M=AT^{0.5}\exp\left(\frac{Ja^2}{k_BL_0^2}\frac{B_s-B}{B_sT}\right). 
\end{equation}
With experimental values of $a_2$ and $J$, we find $L_0$=$0.9a$, which is indeed the physically expected result.

\smallbreak
\noindent \textbf{5. Mott quantum phase transition in cold atomic gases and moiré superlattices.}

\textit{QPT in 2d cold atom Boson systems}

Quantum phase transitions have been studied in several cold atom systems. Data suitable for our analysis were reported in \cite{Chin_2d_ColdAtoms} for a system of 4000-20000 atoms of Cs-133 placed in a 2d square optical lattice with lattice spacing $a$=$\lambda/2$=$0.532$ $\mu$m. The tunneling energy in the lattice is $t$=$k_B\times2.7$ nK. The gas is confined in a wide $2d$ trap with lateral extension of about 80 $\mu$m.  The dependence of an equilibrium atom concentration, $n$, on chemical potential, $\mu$, and temperature was obtained from imaging of the $n(x,y)$ distribution.  The studied transition is between the Mott insulator state with zero occupation number (the vacuum state) and the superfluid state; it is driven by the change of the chemical potential. It is expected that the transition can be explained by the Bose-Hubbard model with critical exponents $\nu$=1/2  and $z$=2. However, apparently there is no exact critical theory that can predict the variation of the scaling function across the transition.

The scaling analysis carried out in \cite{Chin_2d_ColdAtoms} confirmed the values of the exponents. The scaled data traced from this work are shown in Fig. 5a as a blue line, which represents (as in \cite{Chin_2d_ColdAtoms}) the average of the data in the lowest temperatures indicated in the figure. The exponential dependence appears to be a good approximation to the data near the critical point as shown in the figure as a black solid line given by $y$=$a_1 \exp (a_2x)$, with $a_1$=0.095 and $a_2$=1.1, both dimensionless parameters. Notice that the choice of x-axis (in which we followed Ref. \cite{Chin_2d_ColdAtoms}) is different from previous figures.

The analysis of the system is similar to what we used for the Heisenberg chains. Since we have the transition from the vacuum state, at the critical point, the elementary excitation are again Maxwellian particles with typical velocity $v$=$\sqrt{{2k}_BT/m^\ast}$.  Under the tight-binding approximation, the effective mass is related to the tunneling energy as $m^\ast$=$\hbar^2/2ta^2$. As in previously considered clean systems, the dephasing length is set by the ballistic propagation, $L_\varphi$=$v\tau_P$. Also, as before, we are not interested in the prefactor and represent it by a constant multiplied by a temperature to an appropriate power. With this input we have for an occupation number  
\begin{equation}
N=AT \exp\left(\frac{4\ t\ a^2}{\mu_sL_0^2}\frac{\mu-\mu_s}{k_BT}\right).
\end{equation}
It was found in \cite{Chin_2d_ColdAtoms} that $\mu_s$=$-4.6t$. Using this value we find that the experimental seeding scale of the renormalization group is related to the period of the optical lattice as $L_0$=0.9$a$, in agreement with the physical expectations. 

\textit{Mott QPT in semiconductor moiré superlattice}

	Next, we consider Mott transition in MoTe$_2$/WSe$_2$ moiré superlattice studied in \cite{Mak_QPT_Mott}. The density of carriers in this experiment was fixed at half-band filling. The out-of-plane electrical field was used to change the effective interaction strength $U/W$, where $U$ is on-site Coulomb repulsion and $W$ is the bandwidth, thus driving the Mott transition from metallic to insulating state. We have traced $R(T)$ curves from Fig.2 in \cite{Mak_QPT_Mott} and carried out the scaling analysis. Similar to \cite{Mak_QPT_Mott}, we find that the best data collapse occurs when $\nu z \approx0.7$ and the prefactor of the scaling function is set to vary with temperature as $\sim T^{-1.3}$; the scaled data are shown in Fig. 5(b).  The data near the critical field can be approximated with an exponential dependence $y$=$a_1 \exp (-a_2x)$, where $a_1$=3.3$\times{10}^6$  $\Omega$ K$^{1.3}$ and $a_2$=10000 K$^{-1/0.7}$.
	
To analyze the critical regime quantitatively, we take a few notes from the original work: (i) that disorder in the bilayer system is weak, (ii) that the Fermi surface completely collapses at critical field $E_c$ so the chemical potential becomes zero at the transition, and (iii) that the dispersion relation of the carriers is quadratic in the wave vector. All this suggests that the picture of the thermally-excited Maxwellian carriers, which we used for Heisenberg chains and 2d cold gas, may be also applicable to the moiré bilayer. This suggests that $z$=2  and $\nu$=0.35 and using these values, Eq. 3 can be adopted as 
\begin{equation}
R=\frac{R_c}{T^{1.3}}\exp\left(\frac{E-E_c}{E_c}\left(\frac{\hbar\sqrt{2/k_BTm^\ast}}{L_0}\right)^{1/0.35}\right)
\end{equation}

For a MoTe$_2$/WSe$_2$ bilayer, the effective mass $m^\ast$ in the kinetic energy can be reasonably well approximated by the effective mass in the current carrying host layer \cite{Wu_MacDonald,Zhang_Mass_Bilayer}. In the studied system, this layer is MoTe$_2$, for which the effective mass is $m^\ast\approx0.5m$ according to \cite{Mak_QPT_Mott}.  We then find, using the experimental value of $a_2$, that the experimental seeding scale of the QPT in the bilayer is $L_0\approx2.4$ nm. It is about half of the period of moiré superlattice in MoTe$_2$/WSe$_2$, $a\approx5$ nm and thus well follows the trend we see in all systems with the periodic potential. We find the close agreement between $L_0$ and $a$ in MoTe$_2$/WSe$_2$ bilayer to be particularly impressive, since computation of the seeding scale involves a nontrivial combination of fundamental constants, which do not cancel each other.

A peculiar feature of the Mott transition in MoTe$_2$/WSe$_2$ bilayer is the temperature-dependent prefactor of the scaling function. This behavior is different from what one expects from generic form of Eqs. 1 and 3, as well as from what was observed in two-dimensional superconducting \cite{Rogachev_QPT_films} and quantum Hall systems \cite{Rogachev_QPT_QHE}. The scaling with a temperature-dependent separatrix between two regimes of the Mott transition was obtained in numerical modeling using dynamical mean-field theory \cite{Dobro_QPT_Mott_scaling}. Also, according to the phenomenological reasoning of Refs. \cite{Gantmakher_LocDeloc,Gantmakher_SIT_review}, this behavior (albeit in the limited range of parameters) can appear in a system that follows a two-parameter scaling; this method was used for analysis of the MIT in the 2d electron gas of Si MOSFETs in \cite{Knyazev}.  It is really interesting to see if such two-parameter scaling could explain the behavior of MoTe$_2$/WSe$_2$.

\noindent \textbf{6. Summary and conclusions.}

In this section we would like to summarize our results and comment on several conjectures made in our model. Hopefully these conjectures will find support from rigorous theoretical arguments. 

1) In our analysis, we have assumed that scaling functions vary exponentially in the quantum critical regime very close to the critical points of QPTs. While it can be seen just as an approximation which allows us to obtain a simple analytical expression, it is actually quite well followed by the data. In our view, the exponential dependence appears here for a reason, most likely from statistical averaging over different disorder realizations. Let us recall in this context a well-known result: the proper computation of a chain of random scatterers requires averaging over the logarithm of conductance \cite{Andresen_STL_expansion}. It is interesting to note that in experiments, the exponential dependence is more pronounced in 1d and 2d systems.
  
2)  We made a conjecture that the exponential variation of conductivity introduced in the scaling theory of localization can also be extended to quantities like magnetization, spin relaxation rate, and occupation number. Let us mention in this regard the similarity between zero-temperature conductance in the scaling theory of localization and the surface tension term for a magnetic system related to the free energy difference imposed by periodic and antiperiodic boundary conditions (Eq. 8.4 in \cite{Sachdev_Book}.) Establishing this connection more rigorously could perhaps explain the universality of Eq.3.  

3) The central conjecture of our model is the assumption that in interacting systems, the dephasing length is given by the semiclassical propagation of a particle or elementary excitation over Planckian time $\tau_P=\hbar/k_BT$. As our work attests, this approximation provides a fairly accurate description of the quantum critical regime in a large number of systems. Our observations pose the question as to if the semiclassical particles with a life-time given by $\tau_P$ represent real, physical processes happening in the quantum critical regime or if this is just a useful imaginary step leading to the correct prediction of a system response. 
	
4) MIT in Si:P is the only transition that follows the non-interacting version of the model. Let us recall that the used relation, $L_\varphi=(dn/d\mu\ k_BT)^{-1/3}$, does not include the Planckian time and also goes contrary to the common view that $L_\varphi$ for non-interacting systems is set by some inelastic process. Intriguing question is what makes Si:P so special, particularly in contrast to MIT in Si:B, which can be explained within the general framework of interacting systems. 

We hope that extension of our analysis will help to identify microscopic processes leading to QPTs in various systems, perhaps even beyond the realm of equilibrium condensed matter physics, and will serve as a starting point (or as a check point) of microscopic critical theories.  
\smallskip

\noindent \textbf{ Acknowledgements}
 The research was supported by National Science Foundation under awards DMR1904221 and DMR2133014.

\appendix
\section{Scaling theory of localization. Correlation length and its exponent.}   
(Adapted from E. Mishchenko, Advanced Solid State Physics lecture series, University of Utah (2007))

The scaling theory of localization  considers dimensionless conductance $g$ of a hypercube of size $L^d$. It relates $g$ to conductance  $G$ and conductivity $\sigma$ as

\begin{equation}
  \sigma=\frac{G}{L^{d-2}}=\frac{e^2}{\hbar}\frac{g}{L^{d-2}}
\end{equation}

The theory considers the  $T=0$ situation and makes a conjecture that the conductance of a cube of a bigger size $b^dL^d$ is determined only by $b$ and $g\left(L\right), g\left(BL\right)=f\left(b,g\left(L\right)\right)$. In continuous form this is expressed as a statement that logarithmic derivative of $g$ named as $\beta$ is only a function of itself:
\begin{equation}
  \beta\left(g\left(L\right)\right)=\frac{d \ln g\left(L\right)}{d \ln L}.
\end{equation}
    
The theory also makes the second conjecture that $\beta\left(g\right)$ is a monotonic and continuous function. The form of $\beta\left(g\right)$ that interpolates betweem its expected asymptotic behavior at high and low conductance is shown by the blue line in Fig. A1 for a three-dimensional disordered system of non-interacting electrons.     

\begin{figure}[tbph]
\centering
 \includegraphics[width= 0.7\columnwidth]{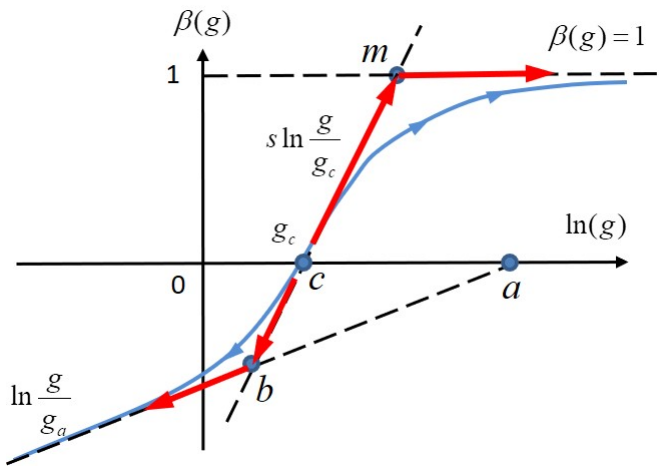}
 \caption{Function $\beta(g)$ versus $\ln g$. Function $\beta(g)$ is shown by the blue line. Asymptotic behavior for low and high conductance and in the critical regime near $g_c$ is indicated with dashed lines. Red arrows indicate the approximating path of the renormalization group flow integration in metallic and insulating regimes. }
 \end{figure}
   
In general, the function $\beta(g)$ is not known. Nevertheless, it is possible to determine the approximate dependences of various physical quantities (conductivity, localization length, etc.) when conductance is close to the critical conductance point, $g_c$.  We use a 3d disordered system of non-interacting electrons as an example.  

At microscopic scale, such a system can be either a metal or an insulator depending on where its conductance falls at the microscopic length scale of an initial seeding cube. For the metal-insulator transition, this microscopic scale is of order of the mean free path $\ell$. The two branches (above and below the horizontal axis) of the function $\beta$ describe what happens to the system when its size is scaled up or, in the language of the renormalization group, determine the flow of the system: 
  
If $g(\ell)>g_c$ , the system “flows” toward the metallic state with asymptotic Ohmic behavior, $\beta(g)=1$.

If $g(\ell)<g_c$, the system “flows” towards the insulator state. Deep in this regime, the conductance is expected to fall off exponentially $g(L)=g_a \exp(-L/\xi)$, which gives the asymptotic behavior $\beta(g)=\ln g /g_a$.

If $g(\ell)=g_c$, the conductance does not change with increasing length scale. This fixed point is unstable; any infinitesimal deviation from $g_c$ leads the system to the metallic or insulating state. 

Let us now find an approximation for conductivity in these regimes.
\bigskip	

(1) $g\left(\ell\right)\equiv g_0>g_c$, \textit{metallic regime:}

For small deviations from $g_c$ we approximate $\beta(g)$ by a linear dependence $\beta=s\ln(g/g_c)$. For large $g$ we approximate $\beta(g)\approx1$. Two lines cross at a point $m$ indicated in the figure; the conductance at the crossing point is $g_m=g_ce^{1/s}$. We now integrate the scaling equation, $(1/\beta) d \ln g=d\ln L$,  along the path indicated by the two red arrows in the top half-plane, from $g_0$ to some $g(L)$ (yet unknown) at a larger $L$

\begin{align}\label{M6}
&\frac{1}{s}\int_{g_0}^{g_m}\frac{d \ln g}{\ln\left(g/g_c\right)}+\int_{g_m}^{g(L)}d\ln g = \frac{1}{s} \ln{\frac{\ln{g_m/g_c}}{\ln{g_0/g_c}}}+\ln{\frac{g\left(L\right)}{g_m}} \nonumber\\
&=\int_{\ell}^{L}{d \ln L}=\ln (L/\ell)
\end{align}
This gives 
\begin{equation}
g(L)=g_m\frac{L}{\ell}\frac{(\ln g_0/g_c)^{1/s}}{(\ln g_m/g_c)^{1/s}}\approx g_c\frac{L}{\ell}\left(se\right)^{1/s}\left(\frac{g_0-g_c}{g_c}\right)^{1/s}
\end{equation}
where the second equality is obtained by using the approximation $\ln\left(g_0/g_c\right)\approx\left(g_0-g_c\right)/g_c\equiv\delta g$ and recalling that $g_m=g_se^{1/s}$. The resultant conductance is proportional to the length of the cube and hence obeys Ohm’s law for $3d$ systems,  $g=\left(\hbar/e^2\right)\sigma L$.	

How big should be the size of the system to start observing this macroscopic Ohmic behavior? By the analogy with the conventional theory of phase transitions this length scale is called the correlation length $\xi$. A natural estimate for this scale is given by the length corresponding to point $m$ in the figure,  $\xi\approx L_m$, where the small-scale variation $\beta\approx s \ln\left(g/g_c\right)$ changes to large scale Ohmic variation $\beta=1$. Thus, it can be found from $g\left(\xi\right)\approx g_m$, which gives $\xi \sim\ell\left(\delta g\right)^{-1/s}$. Next, assuming that the conductance of a “seeding” cube changes linearly with a parameter $y$ that drives the transition, $\left(g_0-g_c\right)/g_c\approx\left(y-y_c\right)/y_c$, we find 
\begin{equation}
\xi\approx\ell\left(\frac{g_0-g_c}{g_c}\right)^{-1/s}\approx \ell \left(\frac{y-y_c}{y_c}\right)^{-\nu},
\end{equation}
This means that correlation length exponent $\nu$  is the inverse of the slope of the scaling function $\beta$ at the critical point, $\nu=1/s$. From Eq. A4, the zero-temperature conductivity on the metallic side of the transition goes to zero as 
\begin{equation}
\sigma=\frac{e^2}{\hbar}\frac{g_c}{\ell}\left(\frac{e}{\nu}\right)^{\nu}\left(\frac{y-y_c}{y_c}\right)^\nu.
\end{equation}
 The coefficient cannot be trusted too much but one can see that the conductivity is controlled by the same correlation length exponent. 

\bigskip
(1) $g\left(\ell\right)\equiv g_0<g_c$, \textit{insulating regime:}

Similar to the metallic regime, we approximate the scaling function $\beta(g)$  with two linear segments, as indicated in Fig. A1 with as the red arrows in the bottom half-plane. Using the notation introduced in the figure, we integrate from $g_0$ towards $g(L)$:
\begin{equation}
\int_{g_0}^{g\left(L\right)}\frac{d\ln{g}}{\beta\left(g\right)}=\nu\int_{g_0}^{g_b}\frac{d\ln{g}}{\ln{g/g_c}}
+\int_{g_b}^{g\left(L\right)}\frac{d\ln{g}}{\ln{g/g_a}}=\frac{L}{\ell},
\end{equation}
which gives 
\begin{equation}
\frac{\ln{g_a/g\left(L\right)}}{\ln{g_a/g_b}}\left(\frac{\ln{g_c/g_b}}{\ln{g_c/g_0}}\right)^\nu=\frac{L}{\ell}\ .
\end{equation}
Noting that $g_a$, $g_b$, $g_c$ are of the same order and approximating $\ln (g_0/g_c) \approx(g_0-g_c)/g_c\approx (y-y_c)/y_c$ we find 
\begin{equation}
g\left(L\right)\approx g_c \exp{\left[-A\frac{L}{\ell}\left(\frac{\left|y-y_c\right|}{y_c}\right)^\nu\right]}
\end{equation}
where $A$  is an unknown constant of order 1. Thus, the conductance is consistent with the localization picture $g(L)\sim \exp (-L/\xi_{loc})$ and has the localization length  
\begin{equation}
\xi_{loc}=\frac{\ell}{A}\frac{y_c^\nu}{\left|y-y_c\right|^\nu}
\end{equation}
The critical exponent $\nu$ is the same as in the metallic regime. The localization length $\xi_{loc}$ diverges at the transition: the closer the system is to $y_c$  the more difficult it is to achieve the localization state, which requires that $\xi_{loc}<L$. 

From the condition that the insulating branch  $\beta\left(g\right)=\ln g/g_a)$ and “critical” branch $( \beta=s\ln g/g_c)$ must intersect, it follows that $s$ must be larger than 1 and that the critical length exponent must be in the interval $0<\nu<1$.

\end{document}